\documentclass[aps,prl,twocolumn]{revtex4}

\usepackage{epsfig}
\usepackage{subfigure}

\begin{document}

\title{Numerical method for integro-differential generalized Langevin and master equations}
\author{Joshua Wilkie}
\affiliation{Department of Chemistry, Simon Fraser University, Burnaby, British Columbia V5A 1S6, Canada}

\date{\today}

\begin{abstract}

We show that integro-differential generalized Langevin and non-Markovian master equations can be transformed into larger sets of
ordinary differential equations. On the basis of this transformation we develop a numerical method for solving such integro-differential equations. Physically motivated example calculations are performed to 
demonstrate the accuracy and convergence of the method.

\end{abstract}

\maketitle

\section{Introduction}

Generalized Langevin equations (GLE) \cite{GLE} and non-Markovian master equations\cite{Numer,Wilk1,Wilk2}, which arise in the treatment of systems interacting with environmental degrees of freedom, often have an integro-differential
form. Unlike ordinary differential equations which can be readily solved using Runge-Kutta, Predictor-Corrector and other well known numerical schemes\cite{NR} there are no general methods for solving equations
of integro-differential type. Here we show that these integro-differential equations can be converted to
ordinary-differential equations at the expense of introducing a new time variable which is treated as if it is of spatial type. [Similar schemes are employed to numerically solve the Schr\"{o}dinger equation for time-dependent Hamiltonians\cite{ttp} and as analytical tools\cite{CG}. There is also some resemblence to
schemes for solving intego-differential equations of viscoelasticity\cite{VE}.] We then develop a numerical
method based on this exact transformation and show that it can be used to accurately solve a variety
of physically motivated examples.

Neglecting inhomogeneous terms resulting from noise, for simplicity, the generalized Langevin equations\cite{GLE} for position $q(t)$ and momentum $p(t)$ of a damped oscillator in one dimension can be expressed in the form
\begin{eqnarray}
dq(t)/dt&=&p(t)/m\label{gle1}\\
dp(t)/dt&=&-m\omega^2q(t)-\int_{-\infty}^t\gamma(t,t')p(t')~dt'\label{gle2}
\end{eqnarray}
where $m$ and $\omega$ are the mass and frequency of the oscillator and 
$\gamma(t,t')$ is the memory function. Defining a space-like time variable $u$ and a function
\begin{eqnarray}
\chi(t,u)=f(u)\int_{-\infty}^t\gamma(t+u,t')p(t')~dt',
\label{GLEA}
\end{eqnarray}
it can be verified by direct substitution that $p(t)$ and $\chi(t,u)$ satisfy the following 
ordinary differential equations
\begin{eqnarray}
dp(t)/dt&=&-m\omega^2q(t)-\chi(t,0) \label{gle3}\\
d\chi(t,u)/dt&=&f(u)\gamma(t+u,t)p(t)+\frac{\partial \chi(t,u)}{\partial u}\nonumber \\
&-&\frac{f'(u)}{f(u)}~\chi(t,u)\label{gle4}.
\end{eqnarray}
Here we have introduced a differentiable damping function $f(u)$ (with $f(0)=1$) which plays a useful role in the numerical scheme
we will introduce to solve the ordinary differential equations (\ref{gle1}), (\ref{gle3}) and (\ref{gle4}). [Note that $f'(u)=df(u)/du$.]

Neglecting inhomogeneous terms, non-Markovian master equations\cite{Numer,Wilk1,Wilk2} can be written
in the form
\begin{equation}
d\rho(t)/dt=-i[H(t),\rho(t)]-\int_{-\infty}^tK(t,t')\rho(t')~dt'
\label{MASTERA}
\end{equation}
where $\rho(t)$ is the time-evolving reduced density matrix of the subsystem, $H(t)$ is an effective 
Hamiltonian, and $K(t,t')$ is a memory operator. [We employ units such that $\hbar=1$.] Defining an operator
\begin{eqnarray}
\chi(t,u)=f(u)\int_{-\infty}^tK(t+u,t')\rho(t')~dt',
\label{SMO}
\end{eqnarray}
it can be verified by direct substitution that $\rho(t)$ and $\chi(t,u)$ satisfy ordinary differential equations
\begin{eqnarray}
d\rho(t)/dt&=&-i[H(t),\rho(t)]-\chi(t,0)\label{master1}\\
d\chi(t,u)/dt&=&f(u)K(t+u,t)\rho(t)+\frac{\partial \chi(t,u)}{\partial u}\nonumber \\
&-&\frac{f'(u)}{f(u)}~\chi(t,u).\label{master2}
\end{eqnarray}
Here $f(u)$ is again a differentiable damping function such that $f(0)=1$. 

Thus, the integro-differential Langevin equations (\ref{gle1})-(\ref{gle2}) can be expressed in the 
ordinary differential forms (\ref{gle1}) and (\ref{gle3})-(\ref{gle4}) and the integro-differential
master equation (\ref{MASTERA}) can be expressed as the ordinary differential equations (\ref{master1})-(\ref{master2}). To exploit these transformed equations as a practical numerical scheme we must discretize
the $u$ variable on a grid of points so that the number of ordinary differential equations is finite. Once this is achieved the ordinary differential equations can be solved using standard 
techniques\cite{NR}. We use an eighth order Runge-Kutta routine\cite{RK} in our calculations. 

To minimize the number of grid points we choose a damping function $f(u)$ which decreases
rapidly with $u$. In the calculations reported here we used $f(u)=e^{-gu^2}$. In practice fewer grid
points are needed for positive $u$ than for negative $u$, and we found that the points $u_j=(-n+l+j)\Delta u$ for $j=1,\dots, n$ worked well when we chose $l={\rm int}(.338 n)$. Here $u_n=l\Delta u$ is the largest 
positive $u$ value. While accurate solutions can be obtained for almost any non-zero value of $g$ we found the most rapid convergence when values were optimized for the type of equation. Hence, $g$ is specified differently below for each
type of equation. To complete the numerical method we need a representation of the partial derivative with
respect to $u$ on the grid. This could be performed via fast fourier transform techniques\cite{NR}. We
chose instead to employ a matrix representation
\begin{equation}
\left(\frac{\partial}{\partial u}\right)_{j,k}=\frac{(-1)^{j-k}}{(j-k)\Delta u}\label{sinc}
\end{equation}
which is known as the sinc-DVR (discrete variable representation)\cite{DVR}. A discrete variable representation (DVR) is a complete set of basis functions, associated with a specific grid of points, in which functions of the variable are diagonal and derivatives have simple matrix representations\cite{DVR}. DVRs are often used in multi-dimensional quantum mechanical scattering theory calculations\cite{DVR}. In the sinc-DVR\cite{DVR}, which is associated with an equidistantly spaced grid on $(-\infty,\infty)$, partial derivatives can thus be evaluated with a sum
\begin{equation}
\left(\frac{\partial X(t,u)}{\partial u}\right)_{u=u_j}=\sum_{k=1}^n\frac{(-1)^{j-k}}{(j-k)\Delta u} X(t,u_k)
\end{equation}
for any function or operator $X(t,u)$. 
In our calculations we chose $\Delta u$ to equal the time interval $\Delta t$ between output from the Runge-Kutta
routine.

We now discuss applications of the above numerical method to specific models.
\begin{figure}[htp]
\caption{Memory functions $W(t)$ plotted against time.}
\epsfig{file=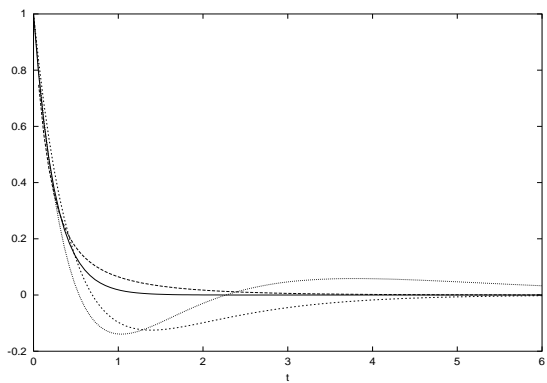,width=3in,height=2.in}
\end{figure}
For the generalized Langevin equation we chose an initial value problem (i.e. $\gamma(t,t')=0$ for $t<t'$ and $\gamma(t,t')=W(t-t')$ for $t\geq t'$) where $W(t)$ has one of the following forms
\begin{eqnarray}
W(t)&=&e^{-4 t}\label{mem1}\\
W(t)&=&\frac{1}{9}\frac{e^{-t}-e^{-10t}}{1-e^{-t}}=\frac{1}{9}\sum_{j=1}^9e^{-j t}\label{mem2}\\
W(t)&=&2e^{-2t}-e^{-t}\label{mem3}\\
W(t)&=&3e^{-2t}-2.8e^{-t}+.8e^{-t/2}\label{mem4}
\end{eqnarray}
which are displayed graphically in Figure 1. The solid curve is (\ref{mem1}), the dashed is (\ref{mem2}), the short-dashed is (\ref{mem3}) and the dotted is 
(\ref{mem4}). These memory functions were chosen to roughly represent the various 
functional forms which can occur physically\cite{GLE} and for ease in obtaining exact solutions. The constants appearing in equations (\ref{gle1}), (\ref{gle3}) and (\ref{gle4}) are chosen as $m=1$ and $\omega^2=10$. Figure 2 shows the functional form
of the exact solutions $q(t)$ (solid curve) and $p(t)$ (dashed), which evolve from initial conditions $q(0)=1$ and $p(0)=.1$, for memory function (\ref{mem1}) over a timescale of 20 units with $\Delta t=.04$. Solutions for the other memory functions (and the same initial conditions) are similar in appearance. These exact solutions were obtained by expoiting the fact that the above memory functions are sums of exponentials (i.e. $W(t)=\sum_{j=1}^{\infty}a_je^{-b_jt}$) from which it follows that one may write
\begin{eqnarray}
dp(t)/dt&=&-m\omega^2q(t)-\sum_{j=1}^{\infty}a_je^{-b_jt}y_j(t)\\
dy_j(t)/dt&=&e^{b_jt}p(t)
\end{eqnarray}
for $j=1,2,\dots$, and solve these ordinary differential equations using standard methods. This 
approach only works for memory functions of this type.
\begin{figure}[htp]
\caption{Position (solid curve) and momentum (dashed) of a damped oscillator.}
\epsfig{file=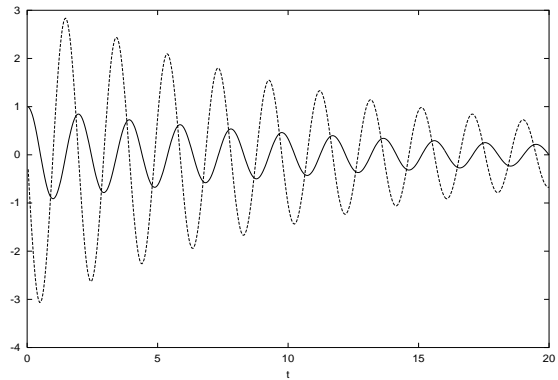,width=3in,height=2.in}
\end{figure}
Approximate solutions were obtained using $g=7/[(n-l)\Delta u]^2$. For negative $u$ we set $W(u)=W(|u|)$.

The negative logarithm of the absolute error in $q(t)$,
\begin{equation}
\epsilon(t)=-\log_{10}|q(t)-q_{{\rm approximate}}(t)|,
\end{equation}
is shown in Figure 3 plotted against time for the values of $n$ indicated in the inset. [The error in $p(t)$ is similar.] As $n$ increases $\epsilon$ increases (on average) and hence the error decreases. The oscillations in $\epsilon$ are caused by periodic intersections of the two solutions. In practice it is impossible to visually distinguish the two solutions when $\epsilon\geq 2$. Note that after a short transient the error (on average) does not increase. This is probably a consequence of the linearity of these equations. Some decline in accuracy with time should be expected when the Langevin equations are non-linear (e.g. a particle in a double-well).
\begin{figure}[htp]
\caption{$\epsilon(t)$ for memory function (\ref{mem1}).}
\epsfig{file=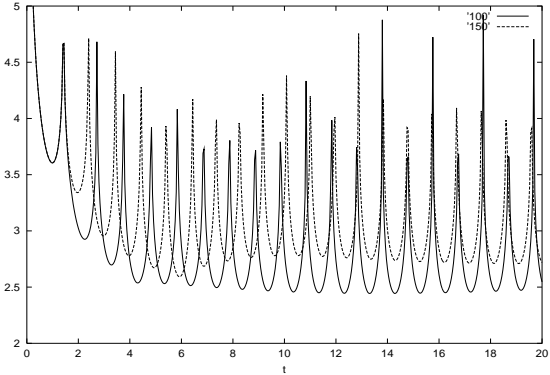,width=3in,height=2.in}
\end{figure}

Figure 4 compares the exact solutions for $q(t)$ (solid curve) and $p(t)$ (short-dashed) with those obtained using our method for $n=150$ (dashed and dotted, respectively) over a time of 40 units. No disagreement is visible. Convergence for memory function (\ref{mem2}) is similar.
\begin{figure}[htp]
\caption{Comparison of exact and approximate position and momentum of a damped oscillator.}
\epsfig{file=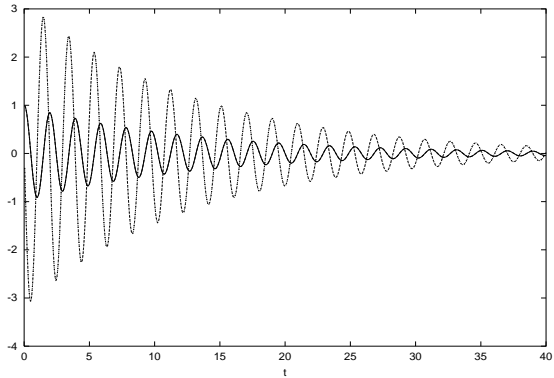,width=3in,height=2.in}
\end{figure}

Memory functions (\ref{mem3}) and (\ref{mem4}) which take negative values and have long time tails require many grid points for convergence. Figure 5 shows
the negative logarithm (base ten) of the absolute error in $q(t)$ for this case. While many grid points are required, high accuracy solutions can clearly be obtained using our method.
\begin{figure}[htp]
\caption{$\epsilon(t)$ for memory function (\ref{mem3}).}
\epsfig{file=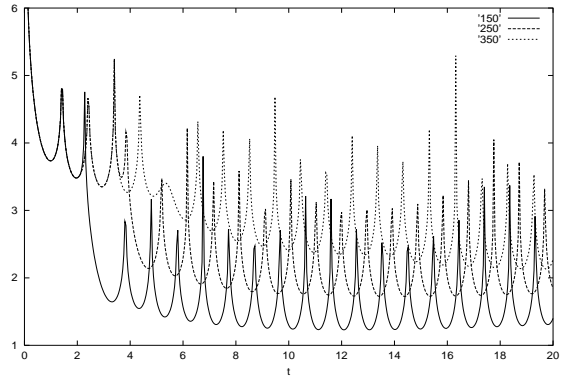,width=3in,height=2.in}
\end{figure}

For the master equation we chose an initial value problem consisting of a dissipative two-level system 
representing a spin interacting with environmental degrees of freedom. If the spin Hamiltonian is $H=\frac{\omega}{2}\sigma_z+\beta\sigma_x$ and the coupling to the environment is proportional to $\sigma_x$ then the equation for the density matrix $\rho(t)$ is of the form\cite{Wilk1,Wilk2}
\begin{eqnarray}
&&\frac{d\rho(t)}{dt}=-i[\frac{\omega}{2}\sigma_z+\beta\sigma_x,\rho(t)]\nonumber \\
&&-C\int_0^t W(t-t')\{\sigma_x^2\rho(t')+\rho(t')\sigma_x^2-2\sigma_x\rho(t')\sigma_x\}~dt'\nonumber\\
&&~~~~
\end{eqnarray}
where the sigmas denote Pauli matrices. Parameters were set as $\omega=1=\beta$ and $C=.2$. We chose to define $\chi(t,u)=\int_0^t W(t-t')\rho(t')~dt'$ which
differs somewhat from the general definition employed in (\ref{SMO}). The transformed equations are
then
\begin{eqnarray}
\frac{d\rho(t)}{dt}&=&-i[\frac{\omega}{2}\sigma_z+\beta\sigma_x,\rho(t)]-2C\{\chi(t,0)\nonumber \\
&-&\sigma_x\chi(t,0)\sigma_x\} \label{stu1}\\
\frac{d\chi(t,u)}{dt}&=&e^{-gu^2}W(u)\rho(t)+\frac{\partial \chi(t,u)}{\partial u}\nonumber \\
&+&2g u~\chi(t,u)\label{stu2}.
\end{eqnarray}
Theory predicts that the memory function $W(t)$ for this problem is approximately gaussian in form\cite{Wilk2}. However, we were unable to obtain an exact solution of the master equation for this case\cite{Note}. Instead we approximate
the gaussian via the similar function $W(t)=14e^{-7.4t}-13e^{-8t}$. Exact solutions for 
\begin{eqnarray}
\langle \sigma_z\rangle (t)&=& {\rm Tr}\{\sigma_z\rho(t)\}=\rho_{11}(t)-\rho_{00}(t)~~({\rm solid-curve}) \\
\langle \sigma_x\rangle (t)&=& {\rm Tr}\{\sigma_x\rho(t)\}=\rho_{10}(t)+\rho_{01}(t)~~({\rm dashed})\\
\langle \sigma_y\rangle (t)&=& {\rm Tr}\{\sigma_y\rho(t)\}=i(\rho_{10}(t)-\rho_{01}(t))~~({\rm short-dashed}) \nonumber \\
\end{eqnarray}
and initial conditions $\langle \sigma_z\rangle(0)=1$ and $\langle \sigma_x\rangle (0)=0=\langle \sigma_y\rangle (0)$ were obtained in the same way as for the generalized Langevin equations and are plotted vs time in Figure 6. 
\begin{figure}[htp]
\caption{Spin $x$ (solid curve), $y$ (dashed) and $z$ (short-dashed) components. }
\epsfig{file=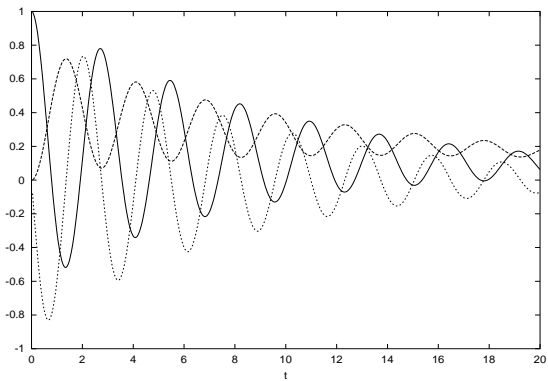,width=3in,height=2.in}
\end{figure}
For the approximate method we used $g=11/[(n-l)\Delta u]^2$ and for negative $u$ we set $W(u)=W(|u|)$. From Figure 7 where we plot 
\begin{equation}
\epsilon(t)=-\log_{10}|\langle \sigma_z\rangle(t)-\langle \sigma_z\rangle_{{\rm approximate}}(t)|
\end{equation}
against time we see that convergence of the numerical method is very rapid for these equations. [Similar accuracies are achieved for $\langle \sigma_x\rangle$ and $\langle \sigma_y\rangle$.] 
\begin{figure}[htp]
\caption{$\epsilon(t)$ for $\langle \sigma_z\rangle$ }
\epsfig{file=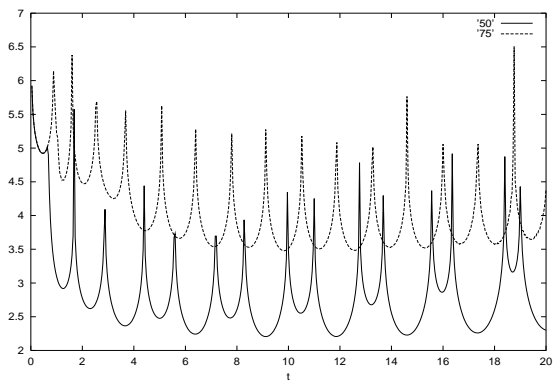,width=3in,height=2.in}
\end{figure}

Thus, we have shown that accurate solutions of integro-differential equations 
can be obtained via transformation to a larger set of ordinary differential
equations. Because this transformation is exact we expect that the method
will also work for equations not considered in this manuscript. It should 
be possible to obtain accurate solutions for such equations via the following steps. First find an approximation of the memory function
or operator which will allow exact solutions to be obtained. Optimize the 
numerical method by finding the best $g$ for the model equations. Finally,
apply the numerical method to the original equations and look for convergence
of the solutions with increasing $n$. 

The author gratefully acknowledges the financial support of the Natural Sciences and Engineering Research Council of Canada.


\begin{thebibliography}{99}

\bibitem{GLE} See for example: G. Frenkel and M. Schwartz, Europhys. Lett. 50, 628 (2000).

\bibitem{Numer} C. Meier and D.J. Tannor, J. Chem. Phys. 111, 3365 (1999).

\bibitem{Wilk1} J. Wilkie, J. Chem. Phys. 114, 7736 (2001).

\bibitem{Wilk2} J. Wilkie, J. Chem. Phys. 115, 10335 (2001).

\bibitem{NR} W.H. Press, S.A. Teukolsky, W.T. Vetterling and B.P. Flannery, {\em Numerical Recipes in Fortran 77, Second Edition}, (Cambridge University Press, Cambridge, 2001).

\bibitem{ttp} U. Peskin and N. Moiseyev, J. Chem. Phys. 99, 4590 (1993); P. Pfeifer and R.D. Levine, J. Chem. Phys. 79, 5512 (1983).

\bibitem{CG} J. Wilkie, Phys. Rev. E 62, 8808 (2000); G. Chen and R. Grimmer, J. Diff. Eqns. 45, 53 (1982).

\bibitem{VE} See the following and references therein: S. Shaw and J.R. Whiteman, Comput. Methods Appl. Mech. Engrg. 150, 397 (1997); R.C.Y. Chin, G. Hedstrom and L. Thigpen, J. Comput. Phys. 54, 18 (1984).

\bibitem{DVR} D.T. Colbert and W.H. Miller, J. Chem. Phys. 96, 1982 (1992).

\bibitem{RK} DOP853.f, E. Hairer and G. Wanner,\\ http://elib.zib.de/pub/elib/hairer-wanner/nonstiff/.

\bibitem{Note} Laplace transforms of the exact solutions are readily obtained
but inverting the transform is problematic.

\end{thebibliography}
\end{document}